**Drone Data Analytics for Measuring Traffic Metrics at Intersections in High-Density Areas**

**Qingwen Pu**
Graduate Research Assistant,
Department of Civil & Environmental Engineering
Old Dominion University (ODU),
4700 Elkhorn Ave, Norfolk, VA 23529, USA
Tel: +1-757-462-6444; E-mail: qpu001@odu.edu

**Yuan Zhu, Ph.D., Corresponding Author**
Professor,
Inner Mongolia Center for Transportation Research, Inner Mongolia University,
Rm A357A, Transportation Building, Inner Mongolia University South Campus,
49 S Xilin Rd, Hohhot, Inner Mongolia, 010020, China
Tel: 86-499-6782; E-mail: zhuyuan@imu.edu.cn

**Junqing Wang**
Graduate Research Assistant,
Department of Electrical and Computer Engineering
Old Dominion University (ODU),
4700 Elkhorn Ave, Norfolk, VA 23529, USA
Tel: 1-(757) 359-3865; E-mail: jwang014@odu.edu

**Hong Yang, Ph.D.**
Associate Professor,
Department of Electrical and Computer Engineering
Old Dominion University (ODU),
4700 Elkhorn Ave, Norfolk, VA 23529, USA
Tel: 1-(757) 683-4529; E-mail: hyang@odu.edu

**Kun Xie, Ph.D.**
Associate Professor,
Department of Civil & Environmental Engineering
Old Dominion University (ODU),
4700 Elkhorn Ave, Norfolk, VA 23529, USA
Tel: +1-757-683-4304; E-mail: kxie@odu.edu

**Shunlai Cui**
Graduate Research Assistant,
School of Transportation and Logistics, Southwest Jiaotong University
Rm 405, Second Teaching Building, Southwest Jiaotong University Xipu Campus,
999 Xian Rd, Chengdu, Sichuan, 611700, China
Tel: 86-15092091526; E-mail: cuishunlai@my.swjtu.edu.cn

**DATA AVAILABILITY STATEMENT**

The High-Density Intersection Dataset is available for download at https://github.com/Qpu523/High-density-Intersection-Dataset.

**ABSTRACT**

This study employed over 100 hours of high-altitude drone video data from eight intersections in Hohhot to generate a unique and extensive dataset encompassing high-density urban road intersections in China. This research has enhanced the YOLOUAV model to enable precise target recognition on unmanned aerial vehicle (UAV) datasets. An automated calibration algorithm is presented to create a functional dataset in high-density traffic flows, which saves human and material resources. This algorithm can capture up to 200 vehicles per frame while accurately tracking over 1 million road users, including cars, buses, and trucks. Moreover, the dataset has recorded over 50,000 complete lane changes. It is the largest publicly available road user trajectories in high-density urban intersections. Furthermore, this paper updates speed and acceleration algorithms based on UAV elevation and implements a UAV offset correction algorithm. A case study demonstrates the usefulness of the proposed methods, showing essential parameters to evaluate intersections and traffic conditions in traffic engineering. The model can track more than 200 vehicles of different types simultaneously in highly dense traffic on an urban intersection in Hohhot, generating heatmaps based on spatial-temporal traffic flow data and locating traffic conflicts by conducting lane change analysis and surrogate measures. With the diverse data and high accuracy of results, this study aims to advance research and development of UAVs in transportation significantly. The High-Density Intersection Dataset is available for download at https://github.com/Qpu523/High-density-Intersection-Dataset.

## INTRODUCTION

Hohhot, one of the most congested cities in China, experiences frequent and dynamic traffic conditions, which heightens the likelihood of traffic conflicts. Expanding traffic infrastructure is often impractical, especially when issues arise due to poor driving habits or inadequate traffic control measures. As a result, traffic engineers have shifted their focus toward understanding how vehicle behavior and traffic management strategies contribute to these problems, seeking solutions that improve safety and efficiency without relying solely on new infrastructure.

While vehicle trajectory and turning movement data can be obtained from various sources, the unique aviation perspective provided by drones offers a valuable alternative for analyzing high-density intersections in congested cities. This study aims to address this flaw by creating a high-altitude drone dataset for high-density urban road intersections. While most existing research combines deep learning (DL) with computer vision (CV) to recognize ground target objects, there is a need for advanced CV models designed explicitly for vehicle recognition from the perspective of drones. This study focuses on vehicles from the viewpoint of drones, improves the YOLO model, and creates a YOLOUAV model for object detection to capture detailed traffic flow features. This approach significantly improves over traditional data collection methods for traffic analysis.

With the rise of unmanned aerial vehicle (UAV) shooting elevations, tracking vehicles has become a significant challenge, particularly in areas with high vehicle volumes and congested intersections, referred to as high-density intersections. The smaller size of target vehicles in such scenarios can decrease accuracy and recognition rate. Additionally, the abundance of targets in existing data sets for UAV vehicle recognition makes it challenging to obtain good detection results using traditional DL techniques. To address these issues, this study involves using a drone flying 120 meters above an urban intersection to capture traffic data, while enhancing the deep learning network structure to improve recognition accuracy. The research also focuses on updating the speed and acceleration algorithm in real-time according to the drone's elevation and correcting the UAV offset caused by pilot adjustments and high-elevation winds. Creating data sets supplies valuable data support and reference for road transportation planning and design in highly congested cities. *conflicts important*

The strong correlation between traffic conflicts and crashes highlights the importance of analyzing traffic conflicts to prevent crashes.(*1*). As such, traffic conflicts derived from aerial video data are an excellent tool for active safety monitoring. This study used detailed vehicle movement data captured in the video to obtain a comprehensive picture of the risk levels at a particular location by quantifying traffic conflicts.

The remainder of the paper is organized as follows: First, the relevant literature is summarized. Next, the methodology for detecting and tracking vehicles in high-density urban intersections is outlined, including the algorithms used to calculate vehicle parameters. A case study at an urban intersection in Hohhot, China is then presented. Finally, the paper concludes by discussing the approach's potential applications in traffic engineering.

## LITERATURE REVIEW

Some studies have contributed to target detection and tracking, with earlier ones mainly focused on identifying specific large targets in close range rather than small targets in long range (2; 3). With the advent of DL, the goal of building general-purpose object detection systems has become more challenging (4), aiming to match the breadth of object detection ability that humans possess. In 2012, the Alexnet (5) was proposed and marked a turning point for target detection with CV, leading to a timeline of milestones in objective detection from 2012 to YOLOv8 (6) in 2023. The classic convolutional network models have shown limitations in recognizing objects from aerial images, particularly vehicles from high-altitude drone perspectives. Consequently, the need for a CV model designed explicitly for higher altitude object recognition has become a crucial challenge in today's advanced technological landscape (7). Consequently, the need for a CV model designed explicitly for recognizing objects from a higher altitude has become a crucial challenge in today's advanced technological landscape. This research proposes a You Only Look Once (YOLO) architecture that performs better on aerial images and demonstrates that changing classes and filters in the three YOLO layers, as well as using Hybrid-SORT (8), can improve accuracy. This work focuses on the application of CV in drones and transportation.

Recent technological advancements have led to the evolution of methods used for collecting traffic data. While traditional devices such as overhead radar sensors and fixed camera systems provide accurate and valuable data (9), they only measure data at specific points, which results in limited information on spatial traffic flow (10). Several datasets have been developed to address this issue, such as the inD Dataset (11), which provides comprehensive data on large-scale urban intersections and natural road user behavior. The AU-AIR Dataset (12) and the CitySim (13) fulfill the criteria for UAVs by incorporating multi-modal data obtained from various aerial sensors. However, these datasets have high GPU performance requirements and limited applicability, making it challenging to use them conveniently. A more portable and practical dataset is needed. The MS COCO DATASET (14) is a large-scale dataset primarily used for object detection and semantic segmentation tasks and includes 2.5 million manually segmented object instances across 91 categories. The KITTI (15) and UA-DETRAC (16) datasets are valuable resources for evaluating environment-aware algorithms and vehicle detection and tracking. However, these datasets (17) may be less effective for target detection in drone views, particularly in high-height, bird-eye views in urban environments focusing on individual vehicles. As one of the most critical components of urban transportation, road intersections (18) are characterized by high traffic volume and fast traffic flow. Consequently, traffic safety issues have become increasingly prominent in this area (*19*). In Hohhot, one of China's most congested cities, traffic congestion at intersections presents a major challenge. A specific high-density urban road intersection dataset, captured by high-altitude drones, enables researchers to study urban traffic patterns in greater detail. This data facilitates the development of strategies to better predict and manage urban traffic congestion (*20*). Researchers from other countries can use this dataset to study similar traffic patterns, adapt strategies, and improve traffic management systems in their urban environments. Therefore, obtaining a dataset focusing more on high-altitude drone imagery of congested urban road intersections is crucial for further advancements in this field.

There are several ways in the UAV-based data collection approach, such as drone video outlier detection and other statistical methods to analyze the intersection for Proactive safety monitoring (10). The studies discussed in this paragraph, while focusing on various traffic scenarios, are not limited to intersections. Representative studies are summarized in Table 1. Gu (*21*) analyzed the collision risk in the highway interchange area by using UAV video data. Chen (*22*) implemented a micro-feature extraction method for mixed traffic flow using UAV data. Wu (*23*) used the masked region convolutional neural network technique to process the traffic videos collected by drones. Ma (*24*) established a traffic conflict prediction model in the highway diversion area by using UAV vehicle trajectory data. Chen (*25*) adopted the kernel trick kernel filter to achieve fast and accurate tracking of vehicles by UAVs. The research above involved extracting alternative safety measures from traffic videos and quantifying traffic conflict risk assessment. Video data recorded the detailed movements of all vehicles, but there was insufficient quantitative data on traffic conflicts to describe risk levels at specific locations. This research addresses this problem fully.

**Table 1 Literatures of computer-vision based traffic study**

| Method | | Literature | Study Area | Content |
|---|---|---|---|---|
| UAV-based Traffic Analys | computer vision | Klautau (2023) (*26*) | Brazil, Para | Simulations were conducted on computer-generated 3D images |
| | | Wang (2023) (*27*) | Guangzhou, China | Drone identification tracking using data augmentation technology |
| | | Di Capua (2023) (*28*) | Naples, Italy | Through drones, integrate augmented reality and virtual reality devices. Alleviate warehouse logistics issues |
| | Traffic Story Risk Analysis | Zhang (2023) (*29*) | Singapore | Predict collision risk in restricted areas of airports |
| | | Gazzea (2023) (*30*) | Bergen, Norway | Estimate traffic flow on the road and predictability in case of extreme weather |
| | | Xing (2023) (*31*) | Beijing, China | The building vulnerability index is generated to provide the spatial distribution characteristics of urban building vulnerability |
| | Calculate Conflict Metrics | Yang (2021) (*10*) | New York, USA | Detection of safety-related anomalies from traffic video data |
| | | Alharbi (2023) (*32*) | Bedford, UK | Combine air traffic flow prediction with intrinsic complexity metrics |
| | | Westhofen (2023) (*33*) | Oldenburg, Germany | Analysis of key indicators for autonomous driving |

The studies discussed above highlight the potential of drones in traffic safety and analysis across various environments. However, there remains a gap in research regarding the use of drones for high-altitude, high-density urban networks and intersection monitoring. This paper addresses this gap by exploring how combining CV with drones can enhance transportation analysis and development.

**METHODOLOGY**

This section presents the vehicle detection and tracking approach, comprising target detection, trajectory tracking, and the measurement of speed, acceleration, and basic turning in a high-density urban intersection. The discussion includes the necessary steps of training data collection, dataset annotation, and weight training, as well as local regulatory constraints and video characteristics. Subsequently, the data extracted from the video is processed and analyzed, encompassing data processing, evaluation of critical parameters in traffic engineering, and measurement of traffic conditions in the urban intersection. The framework of the research is shown in Figure 1.

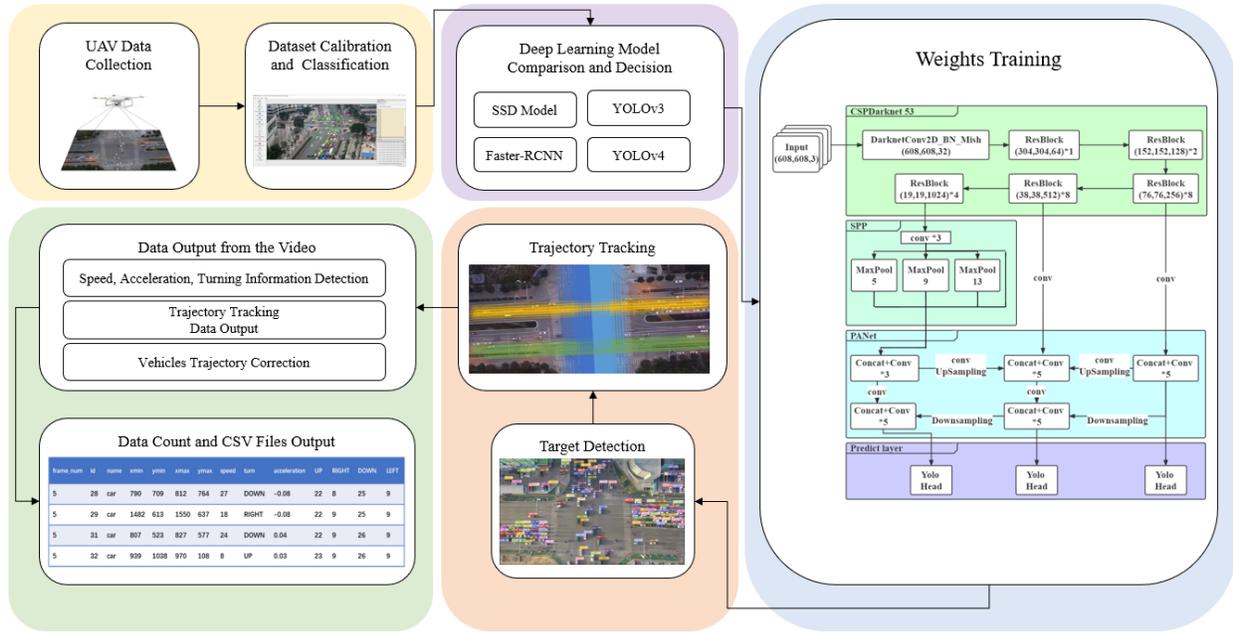

**Figure 1 Method framework diagram**

**Training Data Collection**

The dataset is the cornerstone of target detection and trajectory tracking in DL and CV. It is paramount to procure a suitable dataset and determine the training weights that align with object classification and detection requirements.

The initial task involves identifying and monitoring all vehicles in the high-density urban intersection in the central district of Hohhot, Inner Mongolia, China, where the flying of drones is restricted. Three drones equipped with cameras were deployed to collect information at the intersection. A specific procedure was followed. Initially, a primary UAV model DJI Mavic Air 2 was deployed to record a top-down video of the traffic from 120 meters above the intersection's center. Additionally, two auxiliary drones of DJI Mavic Mini type were positioned at a height of 100 meters and a 30-degree viewing angle in the East-West and North-South arterials, respectively. These drones were used for calibration purposes, as depicted in Figure 2. Each video captured during a flight lasts between 10 and 25 minutes, has a resolution of 1920x1080, and is recorded at 30 frames per second. To ensure diverse data, the videos were shot under various conditions, such as different periods, weather conditions, lighting, and environments. The study area spans 160.6 meters from the left and right edges of the intersection, ensuring comprehensive coverage of the traffic flow. This method obtained over 500GB of videos and generated over 20 million images. It is important to note that local laws and regulations must be strictly followed during filming. The restricted-fly zone imposes a height limit of 120 meters to maintain a safe distance from commercial flight altitude, while flying is strictly prohibited in the no-fly area. Fortunately, with a special permit from the local traffic police detachment, this approach could experiment with filming the main urban sections under 120 meters, overcoming these challenges.

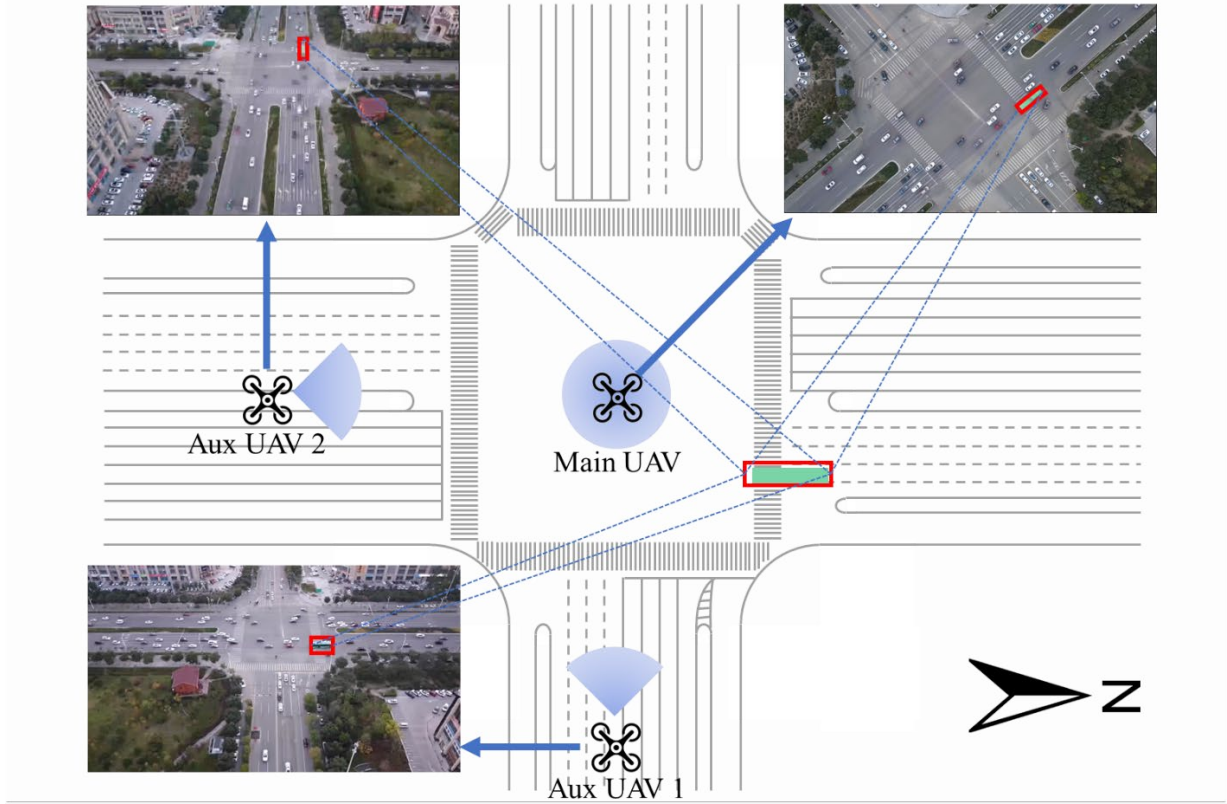

**Figure 2 Three drones with cameras are gathering data at the intersection**

## Dataset Calibration

The UAV images were annotated using LABELIMG (*34*), a graphical image annotation tool. This software allows users to manually select and accurately label vehicles by drawing bounding boxes around them. Each object in the images is categorized into one of three classes: cars, buses, or trucks. The final UAV dataset contains 6,864,000 cars, 353,200 buses, and 156,300 trucks, reflecting the typical distribution of vehicle types in urban intersections.

The number of marked cars is noticeably higher than the two other vehicle categories, which aligns with the typical distribution of vehicle types in urban intersections. Due to truck restrictions within the research area, the truck count is the lowest. Among all the annotated images, the maximum number of annotations is 50, with an average of 46.7. It is important to note that images where objects like telegraph poles or trees obscure the target vehicle should be excluded from the training set to avoid any negative impact.

The next step involves calibrating the dataset, which is traditionally done manually. However, manual calibration is both time-consuming and costly, often prompting organizations to outsource the task to third-party firms. To overcome these challenges, this study introduces a new method for automating the calibration process. This approach significantly reduces the time and expense associated with manual calibration, while maintaining accuracy. To address this challenge, a new method for automatically calibrating datasets has been created. In this study, YOLO format annotation was used, in which the structure of Region of Interest (ROI) is {xcenter, ycenter, width, height}. Unlike this, bbox is usually defined as the upper left corner coordinate $(x_1, y_1)$ and the lower right corner coordinate $(x_2, y_2)$. This study aims to establish a connection between the two, enabling the automatic calibration of the dataset manual data annotation. First, this study uses the current model to track the new video and present the vehicle detection results as a bbox to locate the vehicle's position accurately. Although the calibration box in YOLO format is ROI, this study can convert bbox structure into the form of ROI. The advantage of this approach is that it increases the training set's size without the need for manual data annotation. To ensure data quality, this research introduced quality assurance/quality control (QA/QC) logic. In the automatically generated datasets, this work implemented strict quality checks and excluded any data that did not meet the standards. These QA/QC steps ensure the accuracy and reliability of the final generated data set. After processing a 10-second video, 37,227 vehicles were automatically calibrated. With this approach, this research solves the tedious and expensive problem of manual calibration and ensures the quality of the automatically generated dataset, so that it can be reliably used in research and development in vehicle detection and related fields.

---

**Algorithm 1: Automatically calibrate the core code of the dataset**

---

**Input**: Target detection results $T_{(i)}=\{xmin_{(i)}, ymin_{(i)}, xmax_{(i)}, ymax_{(i)}, i=1...n$

**Output**: $\{xcenter_{(i)}, ycenter_{(i)}, weight_{(i)}, height_{(i)}\}$

1 Initialize the index variable i to 1;

2 Obtain the width and height of the image：

$$size=[picture\ width,\ picture\ height]$$

3 calculate the width and height of each pixel in the image:

$$pixelwidth= 1.0 / (size[0])\ and\ pixelheight = 1.0 / (size[1])$$

4 For each target box i from 1 to n:

5　　Calculate the total length of the target box on the x-axis:

$$totalx_{(i)} = abs(xmax_{(i)} + xmin_{(i)})$$

6　　Calculate the total length of the target box on the y-axis:

$$totaly_{(i)} = abs(ymax_{(i)} + ymin_{(i)})$$

7　　Calculate the difference between adjacent target boxes on the x-axis:

$$differencex_{(i)} = abs(xmax_{(i+1)} - xmin_{(i+1)})$$

8　　Calculate the difference between adjacent target boxes on the y-axis:

$$differencey_{(i)} = abs(ymax_{(i+1)} - ymin_{(i+1)})$$

9　　Calculate the center coordinates of the target box:

$$xcenter_{(i)} = (totalx_{(i)}\ /2.0-1)^*\ pixelwidth$$

$$ycenter_{(i)} = (totaly_{(i)}\ /2.0-1)^*\ pixelheight$$

10　　Calculate the width of the target box:

$$weight_{(i)} = differencex_{(i)}{}^*\ pixelwidth$$

11　　Calculate the height of the target box:

$$height_{(i)} = differencey_{(i)}{}^*\ pixelheight$$

**end**

---

**Model Selection**

There are numerous target detection models in the field of target detection. The models represented by Faster-RCNN (*35*), SSD (*36*), YOLOv3 (*37*), and YOLOv4 (*38*) own the better performance. Scholar Gupta (*39*) compared four models, verifying the better efficiency of YOLOv4 through many experimental analyses. The YOLOv4 model possesses strengths in its specific adaptive capabilities, robust detection, and real-time performance. However, given the diverse application scenarios that necessitate different detection models, it is crucial to determine the most suitable model for detecting videos captured by drones.

In previous studies (*40*), the properties of four models were evaluated based on their training and recognition performance on the COCO dataset. In addition, the performance of these four models was compared using our dataset from drone footage to determine the most suitable model for drone video analysis. Our dataset was fed into four DL models for training and detection at two intersections. The total number of vehicles identified at each instance is displayed on the images (Figure 3).

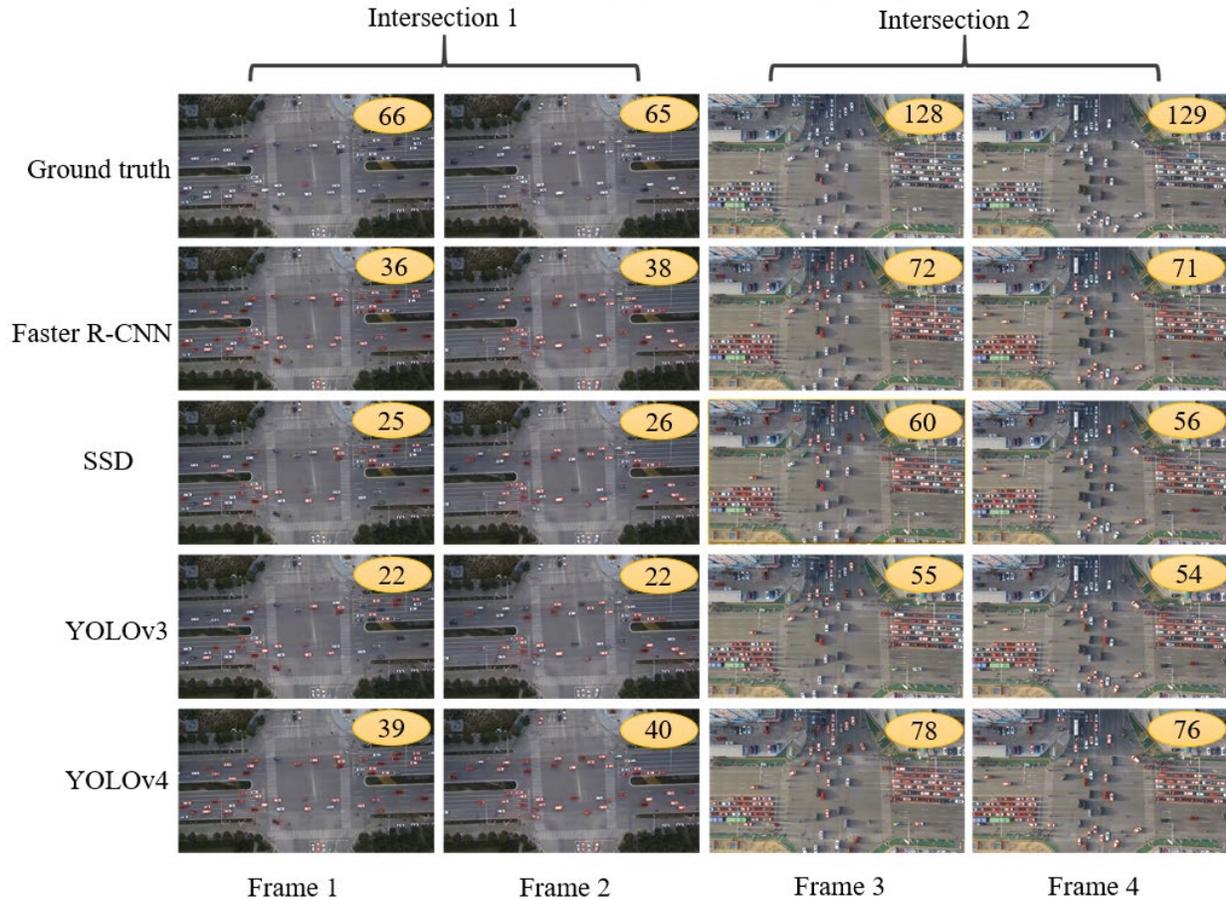

**Figure 3 Experimental results of using the four DL models developed to recognize videos taken by drones**

Based on Figure 3, it is apparent that YOLOv4 surpasses the other models in terms of performance. Despite Faster R-CNN's detection result being comparable to that of YOLOv4, its detection speed is significantly slower, with each detection taking six times longer than YOLOv4. However, for our specific drone dataset, the accuracy of YOLOv4 needed to be improved, necessitating model customization.

## Model Improvement

As stated briefly in the introduction, the DL framework offers an effective approach for training weights in target detection and trajectory tracking applications (*41*). This study constructs the YOLOUAV model based on YOLOv3 and YOLOv4. The improved model selects CSP DarkNet53, which includes 53 convolutional layers, 725 layers × 725 receptive fields, and 27.6 million parameters. The CSPnet architecture splits the stack of the original residuals into two parts: the trunk part, which continues the original stack and a residuals edge that connects directly to the end after some processing. The SPP model is then incorporated by combining it with the convolution of the last feature layer of CSP darknet53. The last feature layer is convolved three times using DarknetConv2D BN Leaky, and the maximum pooled core size is 13×13, 9×9, 5×5, and 1×1. This study proposes Batch Normalization as a solution to the problem of drastic parameter changes in deep neural networks during updates, which can negatively affect gradient stability and speed. By implementing Batch Normalization, the gradient won't vanish due to inappropriate initial parameter settings, and the network's speed can be enhanced. This method avoids early stopping of training. Since the parameters involved in the standardization process include scale and bias, backpropagation mainly calculates the gradient for z, scale, and bias.

$$\frac{\partial loss}{\partial z} = \frac{\partial loss}{\partial z°} * \frac{\partial z°}{\partial z} = scale * \frac{\partial loss}{\partial z°} * \frac{\partial z^{norm}}{\partial z} \tag{1}$$

Among them, $\frac{\partial loss}{\partial z°}$ is the gradient propagation from loss to the input position of the activation function. This step is to calculate each element and put it into the corresponding position of each vector after the calculation, where the m-th element is $\frac{\partial z^{norm}}{\partial z}$.

$$z_n^{norm} = \frac{z_n - Mean(z)}{\sqrt{Var(z)}} \tag{2}$$

Then

$$\frac{\partial z^{norm}}{\partial z} = \frac{\left(1(m = n) - \frac{1}{N}\right)}{\sqrt{Var(z)}} - \frac{\left(z_m - Mean(z)\right)\left(z_n - Mean(z)\right)}{N\sqrt[3]{Var(z)}} \tag{3}$$

The derivative of any $x_n^{norm}$ to any $x^m$ is obtained, and an N × N matrix $\frac{\partial z^{norm}}{\partial z}$ is established, and the (n, m)th element is $\frac{\partial z^{norm}}{\partial z}$.

The derivation for scale is.

$$\frac{\partial loss}{\partial z} = \frac{\partial loss}{\partial z°} * \frac{\partial z°}{\partial scale} \tag{4}$$

For bias, it is calculated together with the neuron's bias in the convolutional neural network, and then accumulated:

$$\frac{\partial loss}{\partial z} = \sum \frac{\partial loss}{bias_m} \tag{5}$$

In addition, the PANet instance segmentation algorithm was utilized in the YOLOUAV model to extract features on multiple occasions. Moreover, the technique of parameter convergence at varying levels was employed to substitute the Feature Pyramid Network.

This study aims to improve the detection and tracking performance in high-elevation images where target pixels are significantly small. To achieve this, the researchers investigate the use of additional data enhancement algorithms in the YOLOUAV model, specifically the Improved Mosaic (*42*) (IM) and Cosine Annealing Decay (CAD) Method (*43*). The original model structure is augmented with these algorithms, and a comparative verification test is conducted to evaluate their efficacy. The study compares the performance of IM and CAD individually and their combination to determine the optimal enhancement approach. The total number of vehicles in the figure is 50.

1. When IM is used in the experiment and CAD is not used, the result is that the loss is equal to 165, and the number of recognized vehicles is 10.

2. When both IM and CAD are used in the experiment, the result is that the loss is equal to 194, and the number of recognized vehicles is 1.

3. When IM and CAD are not used in the experiment at the same time, the result is that the loss is equal to 319, and the number of recognized vehicles is 23.

4. When CAD is used in the experiment and IM is not used, the result is that the loss is equal to 175, and the number of recognized vehicles is 46.

The study's results suggest that the Improved Mosaic falls short in achieving precise detection of small targets in high-altitude captured images. On the other hand, the cosine annealing decay method appears to overcome this issue, as evidenced by its improvement in convergence performance in 50 gradient descent experiments. The hyperparameters adopted for the YOLOUAV model training are presented in Table 2. The training and testing data are captured at different intersections to prevent overfitting and ensure that the model's performance is evaluated accurately.

**Table 2 Hyperparameters in YOLOUAV model**

| Hyperparameter | Value | Description |
| --- | --- | --- |
| Learning Rate | 0.001 (initial) | Gradually reduced using a cosine annealing schedule |
| Batch Size | 64 | Size of mini batches used during training |
| Epochs | 200 | Number of complete passes through the training dataset |
| Image Resolution | 608x608 pixels | Input image size for the model |
| Anchors | Custom calculated | Anchor boxes based on the dimensions of objects in the dataset |
| Optimizer | Adam | Combines advantages of RMSProp and SGD with momentum |
| IoU Threshold | 0.5 | Threshold for classifying positive and negative samples |

**Target Detection**

After feeding the training set into the training network, a weight file with the suffix "weights" can be generated. This weight file is inputted into the YOLOv4 to identify the intersection drone video. Figure 4 (a) shows that the model successfully captures part of the vehicles in the video and identifies specific categories we set. Model 1 has a recognition rate of 22.45%. Model 2, which is Model 1 with modified parameters, has an accuracy of 25.76%, as shown in Figure 4 (b). However, it is evident that modifying only the model parameters does not significantly improve the performance of the model on our dataset.

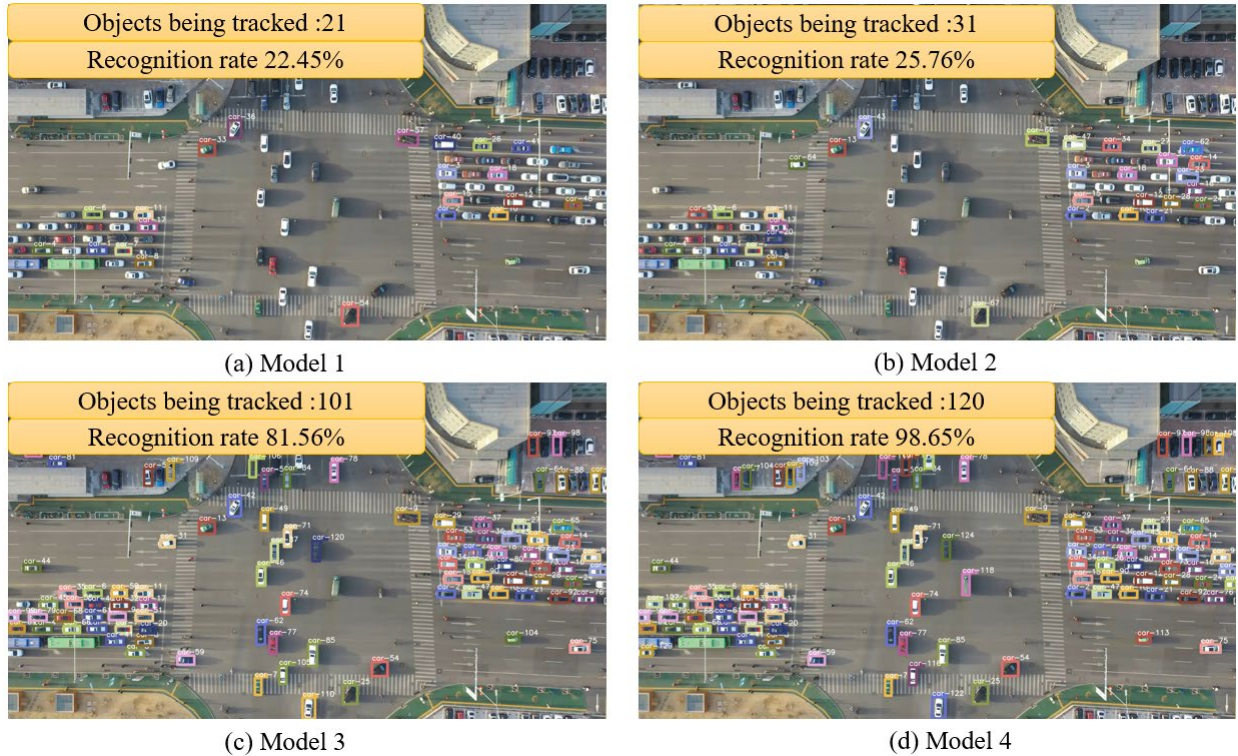

(a) Model 1            (b) Model 2

(c) Model 3            (d) Model 4

**Figure 4 Vehicle identification image**

Models 1 and 2, trained in the above steps, have two obvious weaknesses. One cannot identify each car accurately, and the other is that the frame cannot provide enough stability to track every car shown in the video. Because of the uniqueness of the UAV vehicle recognition, which has the small pixels occupied by vehicles in the image, it leads to failure to extract features and causes difficulty in getting accurate predictions in the next feature prediction layer. This study utilizes the enhanced model discussed in the preceding section as input, modifies the feature extraction layer, and acquires more efficient pixels for the vehicle. During the training process, this research adjusted the parameters by setting max_batches to 6000, steps to 4800 and 5400, and changing the filter to 24 in the three convolution layers preceding the YOLO layer. This led to the continuous training of different models with varying parameters.

Model 4 was trained on an NVIDIA 3060 GPU with 12GB memory. The 17-hour runtime was due to the large size of the UAV dataset and the use of high-resolution images (896x896 pixels). The study aims to ensure accurate vehicle identification under different circumstances, which is why the UAV dataset was added with varying weather and lighting conditions during the later stage of training. The testing dataset used in Figure 4 and Table 3 was collected from UAV video footage at different intersections in Hohhot, China, which were not included in the training set. This ensured that the model's performance was evaluated on unseen data. The experiments have yielded numerous UAV vehicle recognition weights with excellent recognition results, as indicated in Figure 4 (d), where the accuracy has improved from 22.45% to 98.65%.

The resolution of the input feature layer plays a crucial role in improving the feature extraction layer's performance. Although a resolution of $896 \times 896$ yields a slightly better model accuracy, it significantly increases training time and data storage. Therefore, a resolution of $608 \times 608$ is selected as it strikes a balance between accuracy and resource utilization when compared to a resolution of $416 \times 416$.

To demonstrate the superiority and effectiveness of the proposed algorithm, comparative experiments were conducted. The proposed model was compared with several prominent YOLO series models, including YOLOv4, YOLOv5, YOLOv7, and YOLOv8, as well as Faster R-CNN, Cascade R-CNN, RetinaNet, CenterNet, and ATSS. The experimental results are presented in Table 3.

**Table 3 Detection results of various models and the proposed model. (indicates the best result)***

| Models | Precision (%) | Recall (%) | mAP0.5 (%) | mAP0.5:0.95 (%) | Detection Time (ms) |
|---|---|---|---|---|---|
| YOLOv4 | 89.13 | 81.42 | 79.52 | 54.14 | 43 |
| YOLOv5 | 90.57 | 83.63 | 81.92 | 56.87 | 41 |
| YOLOv7 | 93.35 | 86.22 | 85.15 | 60.31 | 38 |
| YOLOv8 | 95.23 | 88.17 | 87.55 | 64.76 | *32 |
| Faster R-CNN | 90.16 | 83.72 | 82.04 | 57.45 | 52 |
| Cascade R-CNN | 91.57 | 85.42 | 83.72 | 59.04 | 50 |
| RetinaNet | 90.05 | 82.84 | 80.93 | 56.17 | 48 |
| YOLOUAV | *98.65 | *89.47 | *88.72 | *65.24 | 39 |

The experimental results in Table 3 show that earlier YOLO series algorithms, such as YOLOv4 and YOLOv5, improve detection accuracy and speed. However, their complex structures and large parameter sizes limit their deployment capabilities on UAV platforms. YOLOv8 optimizes model size and parameter count, significantly enhancing detection performance and speed.

As a two-stage detection algorithm, Faster R-CNN has relatively slower detection speeds and performs less effectively on small objects than the one-stage YOLO series algorithms. Although Cascade R-CNN improves overall detection performance through a multi-stage architecture, it increases computational complexity. While RetinaNet and CenterNet optimize positive and negative sample determination and computational load, their performance in dense scenes and small object occlusions is inferior to the YOLO series algorithms.

Overall, UAV- YOLOUAV demonstrates the best performance across all detection metrics, with superior average detection accuracy and overall detection performance compared to other models. Despite a slight increase in inference time, it can still achieve real-time detection. It indicates that UAV-YOLO excels in precision and recall and has significant advantages in detecting complex scenes and small objects.

**Trajectory Tracking**

In this study, Hybrid-SORT is used for trajectory estimation. This method effectively solves multi-object trajectories by merging weak cues to compensate for strong cues. Compared with DeepSORT(*44*), ByteTrack(*45*), and StrongSORT(*46*), this method analyzes the popular motion-based tracker SORT and achieves superior performance on various benchmarks. Finally, a recognized trajectory was obtained, which outputs the corresponding ID, vehicle category, and coordinates of the bottom left and top right corners of each frame of bbox (xmin, ymin, xmax, ymax). The center point of each vehicle can be calculated using the bbox, and a circle with a radius of one pixel can be drawn in one frame. By processing each frame sequentially, the trajectory of each vehicle is generated and visualized in the video output stream. This study tries to keep the trajectory of every vehicle in every frame. After detecting the vehicle trajectories, an image was generated. Figure 5 shows the complete trajectories of all vehicles in the video.

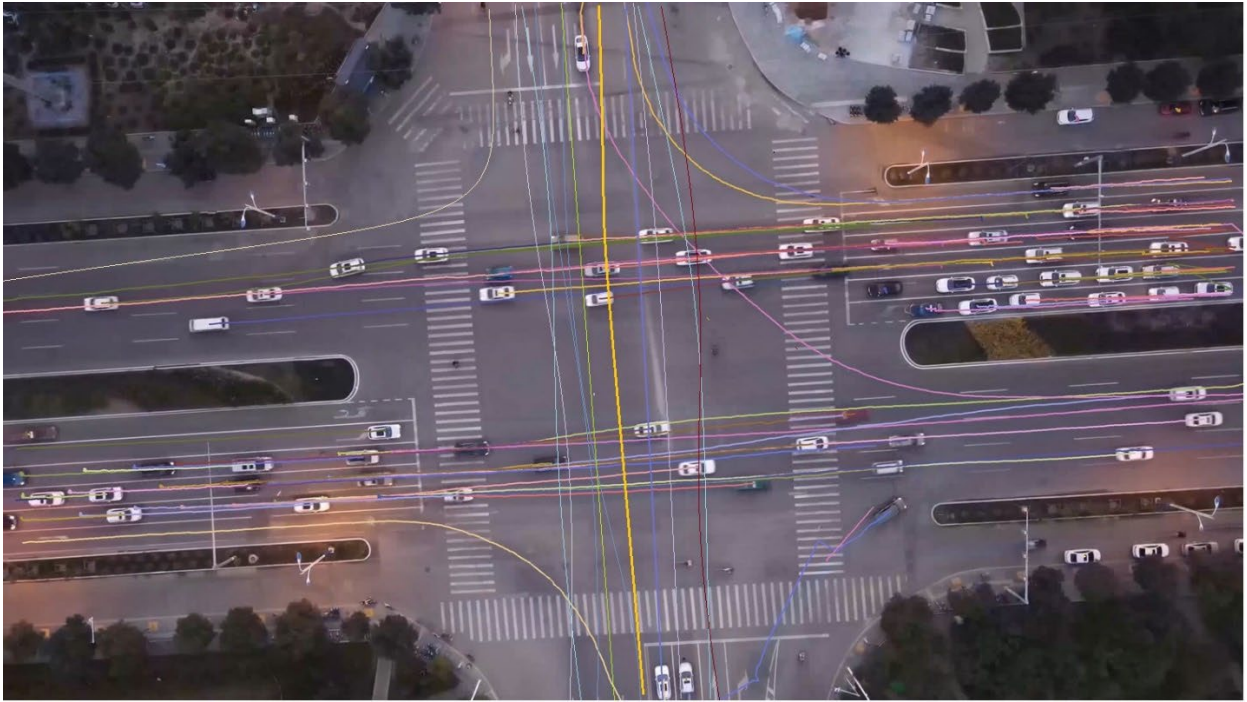

**Figure 5 Vehicle trajectory**

## Speed Detection, Acceleration and Turning

This study aims to obtain more comprehensive traffic parameters by utilizing the output of the previous steps. Measuring the mean speed of vehicles is a critical performance indicator in evaluating traffic conditions; therefore, speed detection is a vital component of this research. YOLOv4 and OpenCV rules were applied to establish a coordinate system in the top left corner of the video. This study repeatedly measured and calibrated the distance between the left and right edges of the video to be 160.6 meters, which is equivalent to 1920 pixels, with each pixel equaling 0.08364 meters in real distance. By knowing the distance per pixel, the speed can be calculated using T(i) and T(i+1), which represent the coordinate values of the bbox of the previous and subsequent frames, respectively. The Euclidean distance between the two frames is primarily used to compute the speed.

| Algorithm 2: Vehicle speed calculation method |
| --- |

**Input**: Target detection results $T_{(i)}$={$xmin_{(i)}$, $ymin_{(i)}$, $xmax_{(i)}$, $ymax_{(i)}$}, i=1...n

$T_{(i+1)}$={$xmin_{(i+1)}$,$ymin_{(i+1)}$, $xmax_{(i+1)}$, $ymax_{(i+1)}$}, i=1...n

**Output**: Vehicle speed $V_{(t+1)}$, i=1...n

1   i=1;

2   **If**   $T_{(i)}$  is the first frame then

3      The output speed is 0

4   **Else**

5      Solve the distance difference in each direction

6      w1 = abs(xmax(i) - xmin(i))

7      h1 = abs(ymax(i) - ymin(i))

8      w2 = abs(xmax$_{(i+1)}$-xmin$_{(i+1)}$)

| 9 | h2 = abs(ymax$_{(i+1)}$ - ymin$_{(i+1)}$) |
|---|---|
| 10 | cx1, cy1 = a1 + w1 / 2, a2 + h1 / 2 |
| 11 | cx2, cy2 = bbox[0] + w2 / 2, bbox[1] + h2 / 2 |
| 12 | Solve for Euclidean distance: |
| 13 | dis = math.sqrt(pow(abs(cx2 - cx1), 2) + pow(abs(cy2 - cy1), 2)) |
| 14 | v = dis * a / 1000 / (1 / 30 / 3600) |
| 15 | **End** |

By calculating the velocity and time difference between frames, the vehicle's acceleration can be determined, which can be used to identify brake and throttle conditions. Additionally, this approach includes the ability to recognize left turns, right turns, and through movements, making it possible to calculate traffic volumes for specific directions.

**Adaptation of Tracking Trajectory for Different Heights**

As drones ascend to higher altitudes, their field of view (FOV) widens, providing more extensive traffic data but at the expense of recognition accuracy. Conversely, lowering the drone's altitude narrows the FOV and enhances observation accuracy. The correlation between altitude and observation accuracy has been extensively researched. By utilizing the speed above detection research, it is feasible to calculate velocity and acceleration based on a specific height and ground cross-sectional length. However, to obtain a more comprehensive traffic video dataset, the drone's flight heights in this study were restricted to between 120 meters and 350 meters. As the drone ascends to greater heights, its perspective expands, and the fixed range solution's speed and acceleration methods mentioned earlier become impractical, as shown in Figure 6 (a).

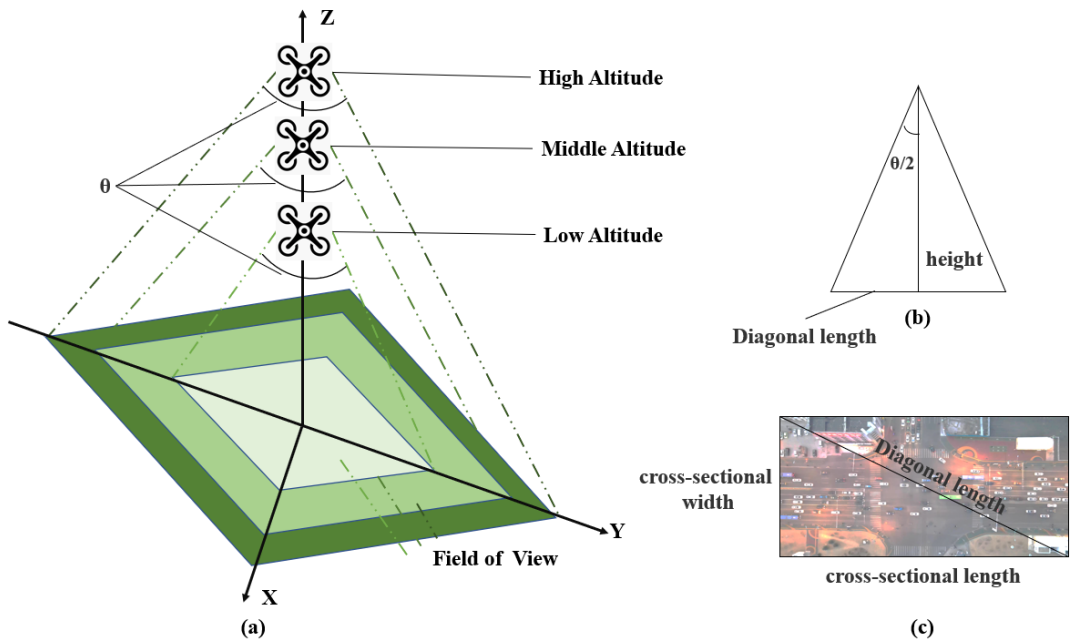

**Figure 6 drone elevation correction algorithm diagram**

This research presents the FOV range for various heights, revealing patterns in the drone field's viewing angle despite elevation differences. The study utilizes the diagonal angle θ of the camera's FOV, converting the drone's altitude problem into an FOV solution, depicted in Figure 6 (b). The Diagonal length is determined by the height automatically measured during the drone's flight, and this study calculates it using the following formula.

$$Diagonal\ length = height * tan\,\theta/2 \qquad (6)$$

As shown in Figure 6 (c), after obtaining the diagonal length, it is known that the pixel aspect ratio is 1920:1080, and then, according to the Pythagorean theorem formula, the ground cross-sectional length can be obtained.

$$length = \sqrt{Diagonal\ length^2 - width^2} \qquad (7)$$

By employing this method, the ground cross-section length of the drone and the velocity and acceleration of the trajectory can be automatically updated according to different drone elevations.

**Drone Trajectory Correction**

This study proposes a method to address the issue of deflection during video recording using a drone. The deflection can be caused by factors such as mis-operation by the drone pilot or interference from environmental conditions like gusts or magnetic fields, which can result in instability of the drone's trajectory output.

To ensure the accuracy of trajectory data, our method includes a detailed three-step process to correct drone camera deflection. Initially, clustering algorithms were deployed to detect camera deflection moments during video recordings. Specifically, the K-means algorithm is utilized to pinpoint the timestamps where deflections occur, which helps to locate distortion periods within the video.

Once a deflection is identified, the deflection angle could be calculated based on the deflected pixel value. This analysis enables the calculation of the deviation from the expected trajectory path. The impact of deflection on data accuracy could be assessed robustly by analyzing these deviations. Finally, the trajectory data is calculated by updating the pixel points in the track output file. Each point is corrected based on its calculated offset, thus enhancing the authenticity of trajectory information. This comprehensive approach ensures the accuracy and reliability of the trajectory data for subsequent analysis.

For instance, Figure 7 (a) displays the image prior to the deflection, while Figure 7 (b) presents the image post-deflection. To provide a more intuitive representation of the deflection scenario, this study extracted the same region from both images and established a coordinate axis. Based on the calculation, the angle between the sidewalk and the coordinate system in Figure 7 (a) is 5°, whereas in Figure 7 (b), the included angle is 0.6°, indicating a significant deflection. Consequently, a correction algorithm was executed for all vehicles, and adjusted trajectories were obtained. To illustrate, consider the range of white vehicles highlighted in Figure 8. Before deflection, this car's bounding box value was {820, 148, 837, 197}. Using the drone trajectory correction algorithm below, it was determined that the bounding box value of this car after deflection is {845, 139, 860, 190}.

| Algorithm 3: drone trajectory correction method |
|---|
| **Input**: Target detection results T(i)= {frame, id, xmin(i), ymin(i), xmax(i), ymax(i), i=1... n |
|     $T_{(i+1)}$={frame, id, xmin$_{(i+1)}$, ymin$_{(i+1)}$, xmax$_{(i+1)}$, ymax$_{(i+1)}$}, i=1... n |
| **Output**:  The output vehicle trajectory |
| 1    Initialize the index variable i to 1; |
| 2 **if**  T(i)  corresponds to the first frame： |
| 3      The output trajectory is empty |

**4 else**

5      Find target pairs with the same ID in both frames

6      **if** $|$ xmin$_{(i)}$- xmin$_{(i+1)}$ $|> 1$:

7        **if** $|$ ymin(i)- ymin$_{(i+1)}$ $| > 1$ then

8          Increment the vehicle count.

11     **if**  the count equals the total vehicle count:

12       Output the Drone Deflection Time

14   **if**  $|$ xmin(i)- xmin$_{(i+1)}$ $|<40$ :

15       Calculate x-direction track offset as: xmin(i)- xmin$_{(i+1)}$

16       Calculate y-direction track offset as: ymin(i)- ymin$_{(i+1)}$

18     Calculate the mean track offset

19     Update the drone's trajectory based on the calculated offsets

**end**

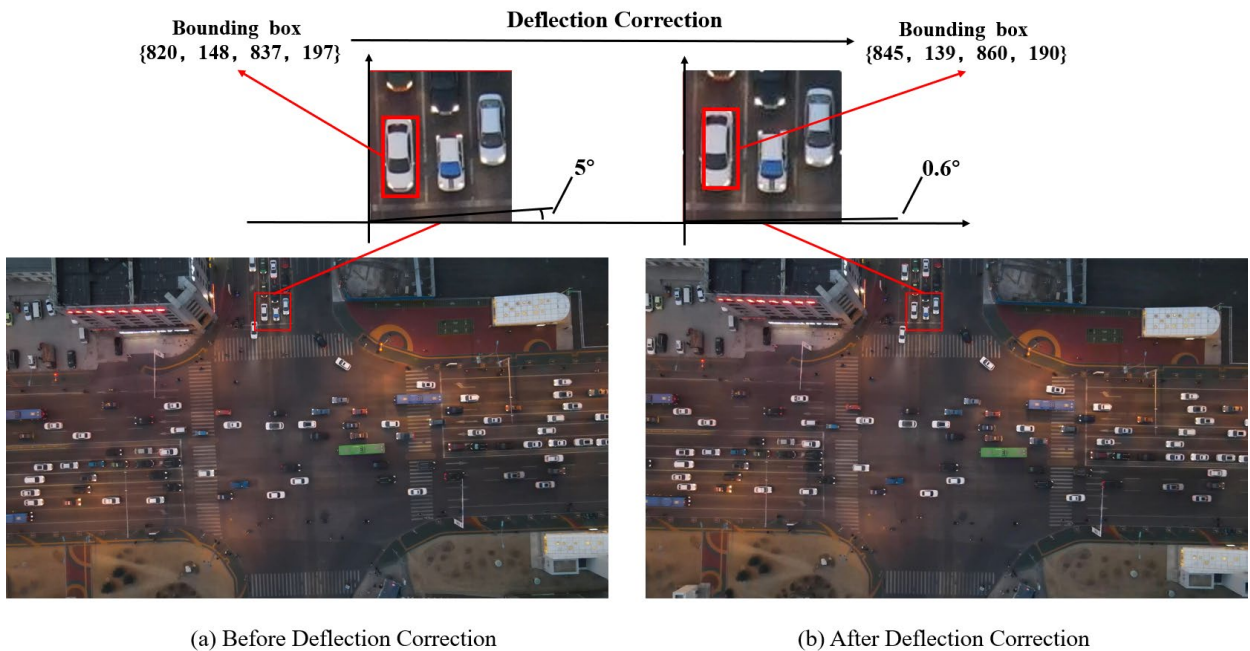

(a) Before Deflection Correction              (b) After Deflection Correction

**Figure 7 Example of UAV deflection correction**

## VALIDATION AND CASE STUDY

### Data Validation

This study generated a high-density urban road intersection dataset using more than 100 hours of video from eight intersections in Hohhot, China, as demonstrated in Figure 8. Five of them (A, B, C, E, and G) are located on Xinhua Street, one of the main arterials in the Hohhot. The other three intersections (D, F, and H) are connected with Xinhua Street. D and H are also located on the city's main roads, S Xing'an Road and S Xilin Road. These intersections are most congested and play essential roles in connecting the city's road network. Each frame can capture up to 200 vehicles simultaneously, accurately tracking over 2 million road users, including buses, trucks, and cars. Furthermore, the dataset records over 50,000 complete lane changes. This dataset is the largest publicly available set of road user trajectories at high-density urban intersections. The study supplies UAV video to make it more useful for tasks like scene classification, road user behavior prediction, or modeling driver behavior. This dataset enables researchers to train advanced machine learning models aimed at improving urban traffic flow predictions and management systems. By leveraging drone-captured data, the dataset provides comprehensive insights into traffic behavior at high-density urban intersections, facilitating the development of intelligent transportation solutions. Moreover, the dataset serves as a crucial resource for urban planners and engineers, offering valuable data to inform the design and optimization of traffic infrastructure in highly congested cities.

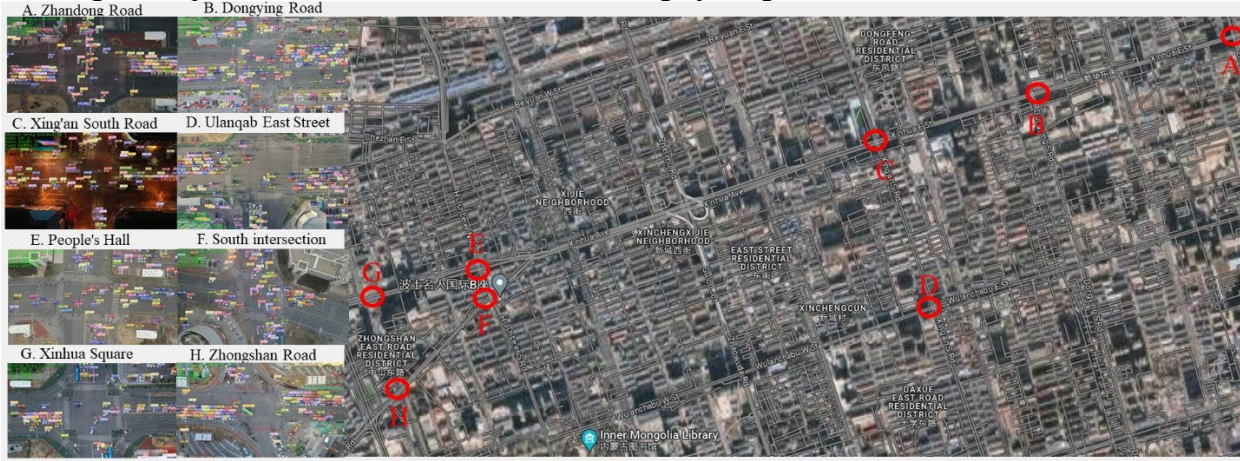

**Figure 8 Distribution Map of the High-Density Intersection Dataset Across Eight Locations**

The approach analyzed the distribution of vehicle turning by traversing all vehicles in each frame, with real-time display in the output video. The findings are summarized in Table 4 including statistics on vehicle turning (xmin, ymin, xmax, ymax), speed, acceleration in the video.

**Table 4 Sample vehicle data extracted from UAV video**

| frame_num | id | name | xmin | ymin | xmax | ymax | Speed (km/h) | acceleration (m/s²) |
|---|---|---|---|---|---|---|---|---|
| 5 | 103 | car | 790 | 709 | 812 | 764 | 27 | -0.08 |
| 5 | 104 | car | 1482 | 613 | 1550 | 637 | 18 | -0.08 |
| 5 | 105 | car | 807 | 523 | 827 | 577 | 24 | 0.04 |
| 5 | 106 | car | 939 | 1038 | 970 | 1080 | 8 | 0.03 |
| 5 | 107 | car | 1847 | 456 | 1913 | 479 | 12 | 0.11 |
| 5 | 108 | car | 771 | 21 | 790 | 72 | 11 | -0.03 |
| 5 | 109 | car | 822 | 1037 | 841 | 1080 | 16 | 0.04 |
| 5 | 110 | car | 505 | 254 | 529 | 319 | 7 | -0.02 |

In Table 4, negative acceleration values represent deceleration, indicating that the vehicles are slowing down.

To validate the accuracy of our speed calculated method, we compared the average speed calculated using the UAV video with the speed derived from this study's vehicle trajectory method. The intersection length used in the video is 160.6 meters. To ensure consistency and comparability, this research selected only vehicles traveling straight from the east-to-west and west-to-east approaches, excluding those making left or right turns.

In the UAV video, the average speed is calculated by dividing the known intersection length by the time a vehicle takes to travel from the entry to the exit of the intersection. The formula used is:

$$v_{\text{video}} = \frac{L}{t} \tag{8}$$

where $L$ is the known intersection length (160.6 meters) and $t$ is the time the vehicle spends traveling through the intersection.

In the vehicle trajectory method, the speed values are obtained from the high-density intersection dataset, which tracks the speed of each vehicle as it moves through the intersection. The average speed for each vehicle is calculated by averaging the recorded speeds over all time steps during its traversal of the intersection. The formula for calculating the average speed is:

$$v_{\text{trajectory}} = \frac{1}{T} \sum_{i=1}^{T} v_i \tag{9}$$

where $v_i$ represents the speed at each time step $i$ and $T$ is the total number of time steps as the vehicle moves through the intersection. This method ensures a more detailed representation of vehicle movement by considering the instantaneous speeds at each time point.

In the vehicle trajectory method, we calculated the speed at each moment as the vehicle moved through the intersection. These speeds were derived from the high-density intersection dataset, generated through the vehicle trajectory tracking process. The average speed for each vehicle was then computed from the trajectory data, providing an accurate representation of the vehicle's movement.

To validate the accuracy of the data in our dataset, this study compared the results of the two methods. we computed the Mean Absolute Percentage Error (MAPE), using the following formula:

$$\text{MAPE} = \frac{1}{N} \sum_{i=1}^{N} \left| \frac{v_{\text{trajectory},i} - v_{\text{video},i}}{v_{\text{video},i}} \right| \tag{10}$$

where $v_{\text{video},i}$ represents the average speed calculated from the video, and $v_{\text{trajectory},i}$ represents the average speed derived from the trajectory data. This research selected data from 198 vehicles, and the resulting MAPE value was 5.12%, indicating a relatively small error between the two methods.

Although MAPE gives us a good understanding of the relative error, it does not provide information about the absolute difference between the speeds calculated by the two methods. To address this, this study also calculated the Root Mean Square Error (RMSE), using the following formula:

$$\text{RMSE} = \sqrt{\frac{1}{N} \sum_{i=1}^{N} \left( v_{\text{trajectory},i} - v_{\text{video},i} \right)^2} \tag{11}$$

The RMSE value was approximately 1.78, indicating that the absolute difference in speed between the two methods is also small.

These two metrics demonstrate that the vehicle trajectory method is consistent with the UAV video method, providing reliable speed estimates. The relatively low error values confirm that the data extracted from the UAV video and the trajectory tracking process are reliable for estimating vehicle speeds in high-density traffic conditions (*47; 48*). Additionally, these results demonstrate the accuracy of our UAV trajectory extraction method and the correctness of the High-Density Intersection Dataset.
.

## CASE STUDY: TRAFFIC DATA COLLECTION

This section presents a case study conducted in the CBD area of Hohhot, China, focusing on the Art College intersection of XINHUA street and XING'AN Road. The study collected driving information for each vehicle and calculated their specific turning information in each frame. The partial schematic diagram of the intersection is depicted in Figure 9 (a), while the UAV data was used to obtain the overall operation of the intersection in this study. Figure 9 (b), (c), and (d) display the average travel times, speeds, and traffic flow for each lane.

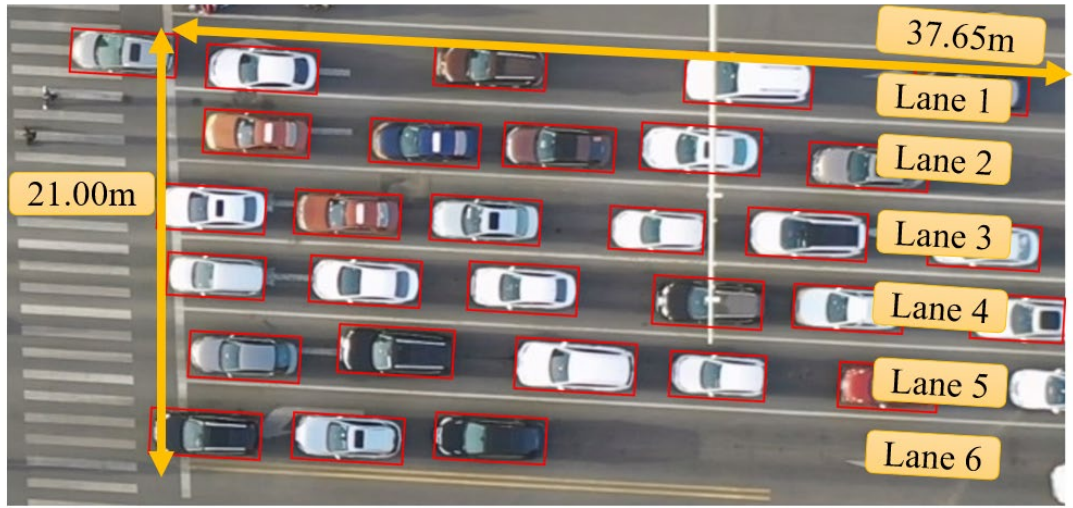

**(a)**

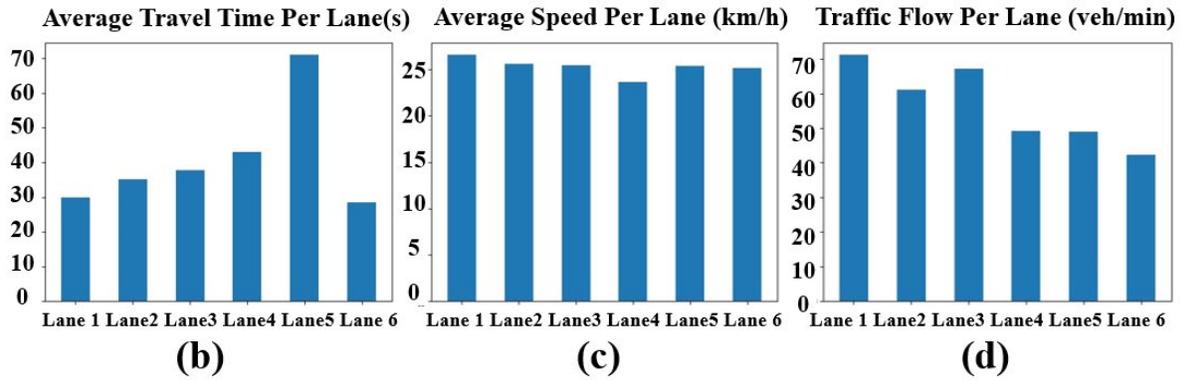

**(b)**                     **(c)**                     **(d)**

**Figure 9 (a) Schematic diagram of actual road with road structure and traffic direction.  (b)(c)(d) Traffic metrics observed under drone monitoring**

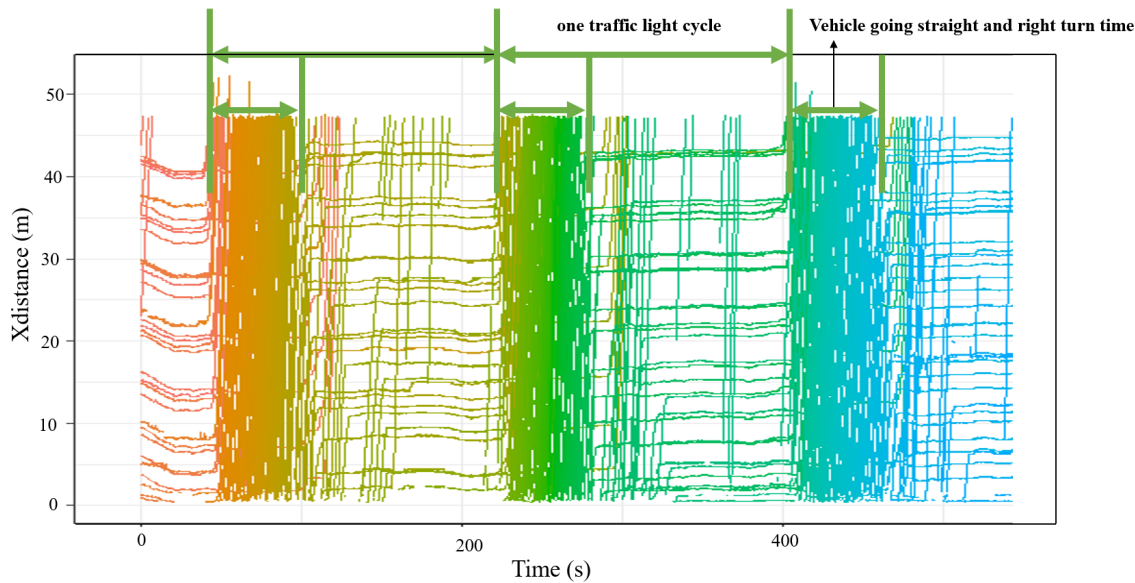

**Figure 10 Time-space diagram of westbound vehicles near the intersection of Xinhua Avenue and Xing'an Road**

Figure 10 displays the time-space diagram of trajectories over a specific time, depicting the traffic flow and signal information of a signalized intersection. The diagram offers additional insights, such as the cycle length of the traffic light and the travel time of vehicles. Additionally, it can portray the queue progression of drivers' reaction time.

Additionally, a graph in Figure 11 (a) maps the y-axis offset of the vehicle, which enables the identification of lane changes based on the cutting lines among the offsets. These lane change movements are then extracted to Figure 11 (b), which displays the vehicle's lane change time and spot speed during the lane change in a box. This study ultimately quantifies the data of lane changes, such as for vehicle id 275 shown in red in Figure 11 (b), where the initial lane was the third lane and the lane change occurred in 242.44 seconds, with a spot speed of 14km/h. After changing lanes twice, the vehicle arrived at the fifth lane. The detection and analysis of lane changes can provide traffic authorities and researchers with valuable insights for finding reckless driving behaviors on certain roadways.

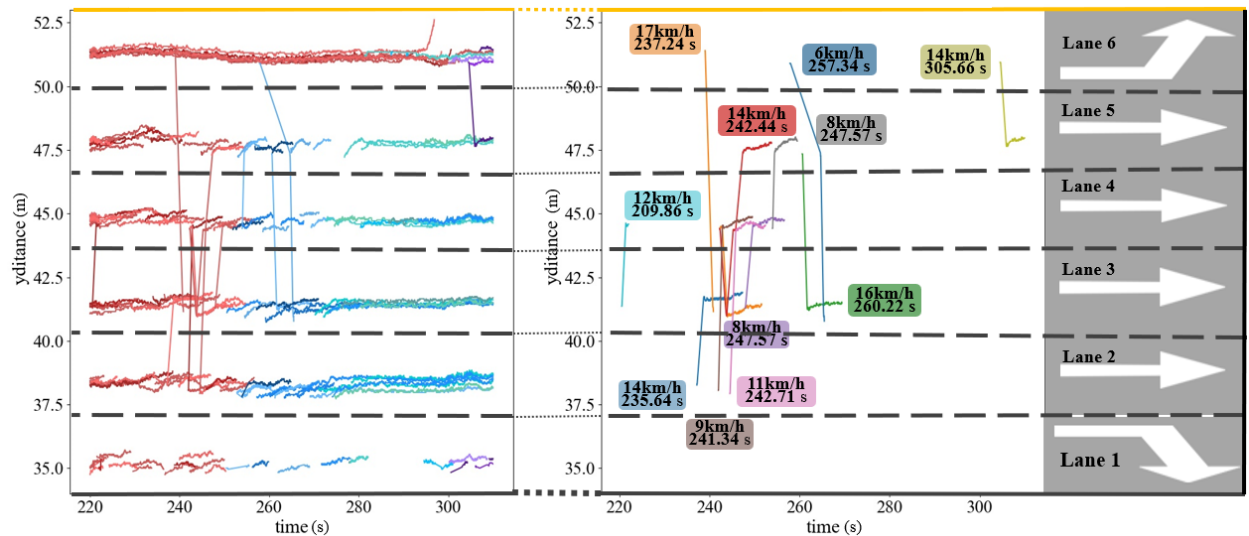

**Figure 11 Vehicle Lane change diagram**

Traditional road safety research primarily uses statistical analysis of vehicle data to investigate traffic crashes. However, Surrogate Safety Measures (SSM) offer an alternative approach to assessing the risk of traffic behaviors that may lead to collisions. When two vehicles interact, one may need to take evasive action to avoid a collision. Surrogate parameters such as Time to Collision (TTC), Post Encroachment Time (PET), deceleration rate, maximum speed, and speed difference are commonly used to measure the severity of conflict or potential collision. Specifically, TTC, PET, and deceleration rate can be used to gauge the severity of the conflict, while maximum speed and speed difference can be used to assess the severity of a potential collision.

$$TTC_f = \frac{D_{i-1(t)} - D_{i(t)} - L_i}{V_{i(t)} - V_{i-1(t)}} \tag{12}$$

Where D is the vehicle position, L is the vehicle length, $V_{i(t)}$ is the speed of the following vehicle, and $V_{i-1(t)}$ is the speed of the preceding vehicle.

$$PET_i = t_{i-1} - t_i \tag{13}$$

Where $PET_i$ is the time difference between the i th vehicle and the following vehicle (that is, the i-1 vehicle) reaching a specified section.

Using the method described above, this study examines the vehicles in each lane, determining the threshold condition for collision avoidance of the following car. Specifically, if the distance between the rear car (i+1 th vehicle) and the main vehicle (i th vehicle) is less than the minimum warning distance $s_{i+1}$ (min), and the maximum deceleration of vehicle i+1 cannot reach the expected deceleration value, a collision could occur. Based on scientific statistics (*49*), a vehicle warning given 2.5 seconds in advance allows for adequate human reaction time and braking distance to stop the car. The vehicle in front should then cooperate by accelerating to avoid a collision.

This study analyzes conflicts between vehicles using a 2-second (2000 ms) value, where a lower TTC value indicates a higher probability of an accident. Figure 12 presents a partial statistical analysis of collision times (ms), revealing scattered values except for potential conflicts with constant values (ConflictID 108) throughout the TTC. Additionally, potential conflicts with more observations (ConflictID 14, 59, and 93) pose a longer risk exposure and greater danger. The study area exhibits a high density of TTC conflicts, particularly at intersections. In conclusion, parking a car too close to a crosswalk or another car behind can result in potentially severe consequences.

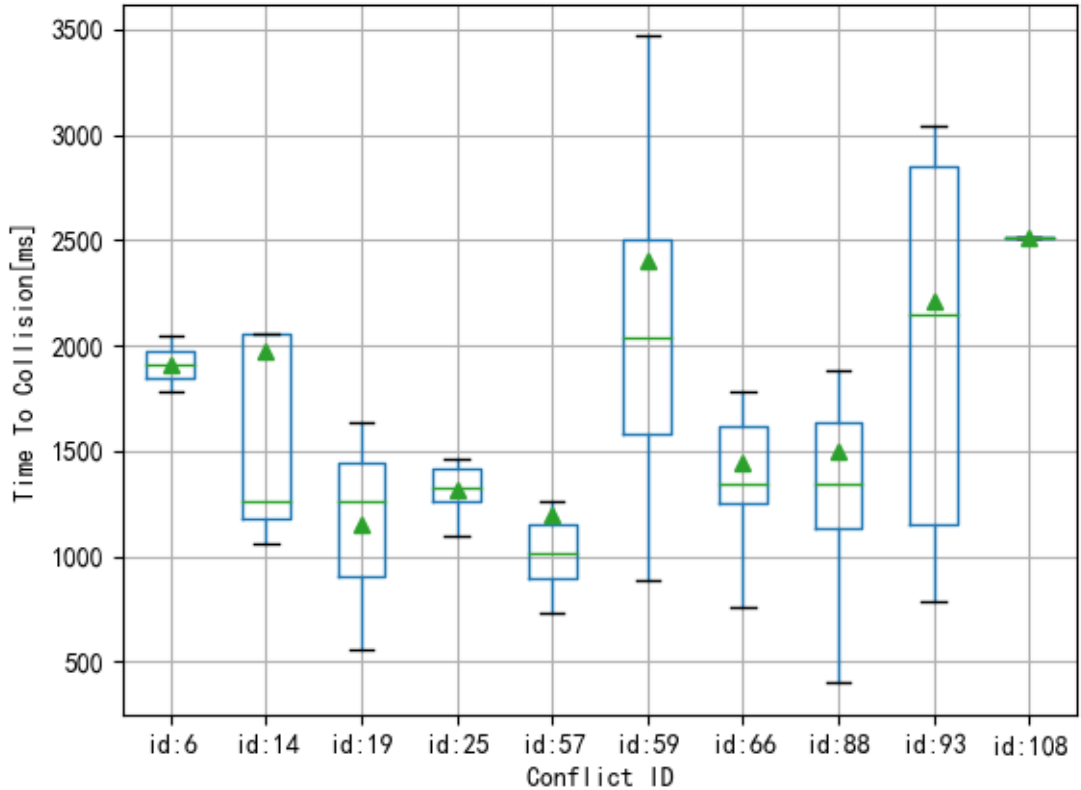

**Figure 12 Distribution of TTC by Conflict ID**

Based on a literature review (*50*) and considering potential conflicts, lower PET (ms) values suggest an increased likelihood of collision. For this study, a proper threshold is 1500 ms, which identifies 356 potential conflicts. Table 5 displays some of the outcomes, with 95 potential PET collisions (26.68% of the total) having a value of less than 1000 ms, indicating a likelihood of collision. This study concludes that most PET conflicts happen on crosswalks, implying that drivers are aggressive and unwilling to yield to other drivers. This scenario is potentially dangerous, as the leading vehicle may abruptly stop, resulting in rear-end collisions.

**Table 5 PET part data and the vehicle safety status**

| PET part data | | | | | | | | |
|---|---|---|---|---|---|---|---|---|
| **Conflict id** | 6 | 14 | 19 | 25 | 57 | 59 | 66 | 88 |
| **PET (ms)** | 1007 | 987 | 601 | 655 | 823 | 1505 | 940 | 897 |
| **The vehicle safety status** | | | | | | | | |
| **id** | speed | acc | time | distance | Speed-difference | collision | collisionID | TTC |
| **2** | 0 | 0 | 9.0 | 35.7 | 31.1 | FALSE | | 4129.1 |
| **19** | 9.91 | 0.1 | 9.0 | 1.4 | 8.3 | TRUE | 14 | 601.4 |
| **59** | 5.75 | -0.02 | 9.0 | 1.9 | 2.9 | FALSE | | 2347.7 |
| **65** | 8.13 | 0.03 | 9.0 | 1.5 | 1.7 | FALSE | | 3045.0 |

After conducting video observation, it has been confirmed that the three marked vehicles pose potential conflict risks. As shown in Figure 13, the locations of potential conflicts at intersections reveal that nearly all potential crashes happen near the crosswalk. This implies that most users follow traffic rules, although some drivers behave aggressively. Notably, three potential collisions occurred outside the crosswalk area involving vehicle id 14, vehicle id 19, and vehicle id 25. These collisions were characterized by three collision times and three post-encroachment times.

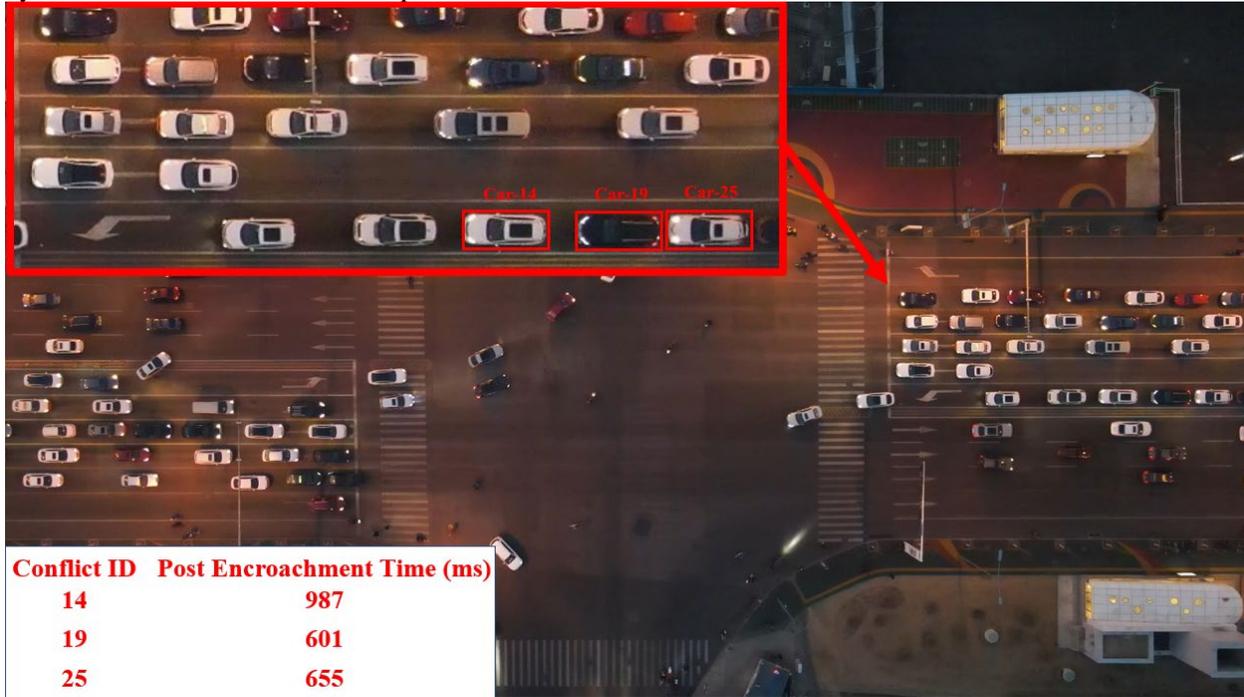

| Conflict ID | Post Encroachment Time (ms) |
|---|---|
| 14 | 987 |
| 19 | 601 |
| 25 | 655 |

**Figure 13 Sample case of conflicts**

The vehicle safety status is obtained, as shown in Table 5. For example, the car with id 19 has a 3.91km/h speed. Its acceleration is 0.1 m/s². It's at 9.033333 seconds of the video. The front car id of this car is 45. The distance between it and the vehicle in front is 1.402625m. The speed difference between it and the car in front is 8.39 km/h, TTC is 601.4475ms, the vehicle id that this car may collide with is 14.

This study found that geometric representation methods can be used to quantify traffic conflicts. However, the dispersed conflict analysis based on centroids has limitations in detecting some critical safety events. A bounding box-based measurement method is more effective in identifying critical conflict events than the centroid-based one. Based on these results, the study collected significant TTC data for each conflict point and calculated the mean TTC value. The study also developed a color-coded system to be the TTC value, with red indicating the highest risk of collision. The period from 0 to 1000 milliseconds was identified as the most likely time frame for potential vehicle conflicts. As a result, this interval was appointed as the initial level and highlighted with a warning red color. The criteria for levels 2 to 5 were then deduced from this data. The results are presented in Figure 14.

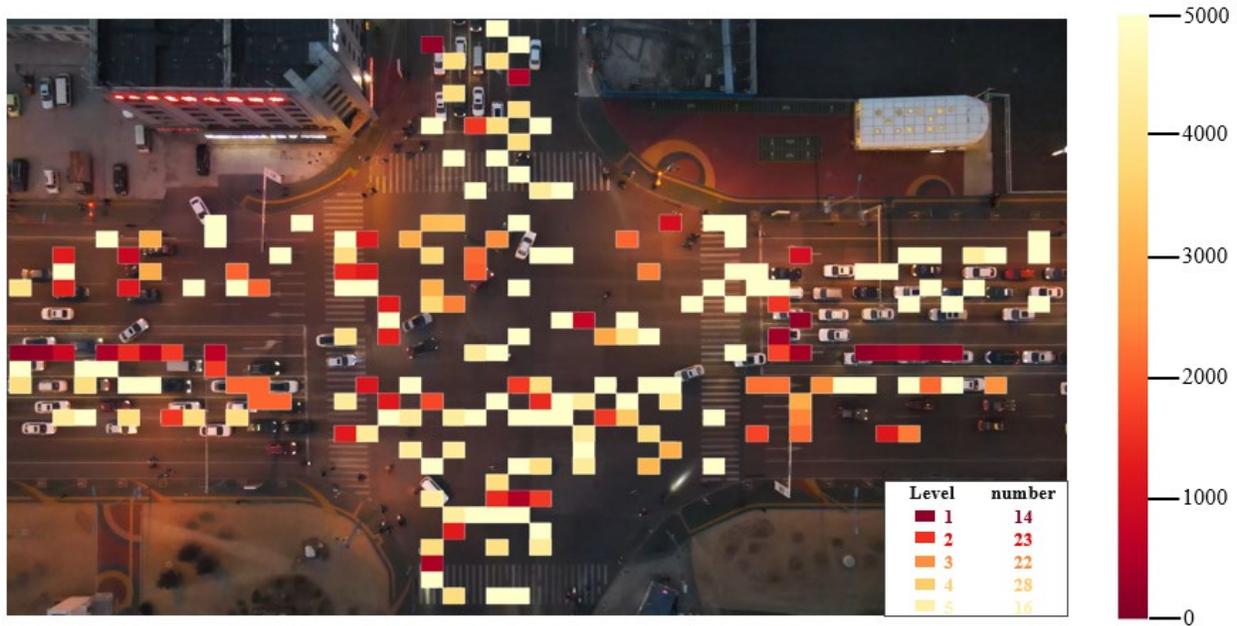

**Figure 14 Heat map of the distribution of critical conflict events at intersections.**

The study found 103 conflict points at the intersection, with left-turning and right-turning vehicles being the primary cause. This aligns with the intersection's actual operation. The accompanying figure indicates that west entrance vehicles illegally turn around on the solid line. During the observation period, 2,197 vehicles entered the intersection, including 2,156 cars (98.13%), 16 buses (0.73%), and 25 trucks (1.14%). Given these findings, the study recommends that the local transportation department address the conflict points by adding more traffic guidance lines and waiting areas to mitigate traffic conflicts.

To validate the accuracy and reliability of the drone data, we conducted a comparative analysis with existing literature on traffic patterns and violations at intersections. Specifically, our findings, such as the high occurrence of illegal left turns and the identification of traffic conflict points, align with those presented by Fu and Liu (*51*), who explored traffic violations at signalized intersections in Hohhot, China.

Their research highlighted that left-turn lane ratios significantly increased traffic violations such as Red-Light Running and Driving in the Inaccurate Oriented Lane. This finding is consistent with the results of our UAV-based traffic analysis, where conflict points predominantly occurred due to left-turning vehicles, as shown in the heat map generated from our drone data. This consistency with previously published research provides a level of validation for our findings, supporting the efficacy of our drone data analytics methodology in identifying critical traffic behavior and conflict points at intersections.

Thus, using a rational model to evaluate traffic safety is crucial in devising effective measures to enhance traffic safety. The research aims to find places with frequent serious traffic issues and create a map showing these intersection conflicts. This map helps pinpoint problem areas, allowing for effective solutions.

To reflect the reality of traffic at the intersection between 5 PM and 7 PM for a two-hour period, the data was smoothed using the expansion method to estimate the number of missing vehicles counts during other periods. The study evaluated vehicle counts every 15 minutes between 5:30 PM and 6 PM, and from 6 PM to 6:45 PM. Once full data was collected, the study calculated the basic parameters of traffic engineering, including a Peak Hour Factor of 0.8428, a Peak Period Factor of 0.8187, a time mean speed of 14.5 km/h, a space mean speed of 12.89 km/h, a density of 41 pcu/km, a capacity of 1494 pcu/h for the south-north direction, a capacity of 1838 pcu/h for the west-east direction, and a safety level of service rating of "Forth".

## CONCLUSIONS AND FUTURE WORK

This study presents a novel approach for detecting and tracking vehicles using DL and CV. The improved YOLOUAV model constructed forms the foundation of this study, which can effectively detect vehicles at high-altitude and high-density intersections. The high altitude and large-scale dataset proposed in this study focuses on recognizing vehicles in high-density urban intersections in China, which is a challenge existing datasets fail to address. If someone is interested in this data set, it can be obtained by contacting the author. Notably, ground recognition differs from recognition at higher elevations, which is more challenging. This technology supplies a convenient solution for recognizing vehicles from UAVs, specifically in high-density traffic urban intersection scenarios. This application is currently underrepresented in research, and this study addresses the issues of speed and acceleration changes caused by UAV height variations and trajectory changes due to UAV offsets. A case study confirms the approach's effectiveness, and traffic engineering parameters are calculated to evaluate the intersection, including traffic flow time and space diagram analysis, vehicle lane change analysis, TTC and PET analysis, and traffic conflict point heat map, which can quickly identify traffic hazards. Based on the road conditions, rational optimization of traffic organization can be achieved.

This proposed methodology has been validated using recorded video datasets, with a MAPE of 5.12% and a RMSE of 1.78, confirming the accuracy of the speed estimation process. These results demonstrate the reliability of the method for real-time traffic monitoring applications. This adaptability allows the system to potentially operate under dynamic conditions, providing immediate insights into traffic patterns and conflicts. Implementing this methodology in real-time involves optimizing the algorithm for faster data processing and ensuring robust data transmission capabilities. Such enhancements will enable transportation professionals to utilize UAV-based data for proactive traffic management and incident response, making it a valuable tool for historical data analysis, real-time traffic control, and safety monitoring.

Based on the available data, passenger cars constitute over 90% of the total vehicles, highlighting the need to explore the potential of non-motor vehicles with drones. Future research should analyze traffic congestion in the leading urban network using traffic simulation and statistical models. Such research would enable real-time monitoring and display of critical traffic parameters such as service level, capacity loss, and travel time index on city roads. Furthermore, the data captured by this study can be used to calculate several metrics in traffic engineering, such as headway, the gap between each vehicle, and the angle when turning, which are difficult to measure using traditional methods. These metrics can help assess potential traffic safety risks and enable proactive safety monitoring. When combined with drone data, this approach becomes a crucial project for collaborating with local traffic police to address traffic issues and optimize conditions. This paper highlights the significance of this method for addressing traffic-related problems.

While this study successfully identified traffic conflict points and illegal driving behaviors using drone data, the absence of ground truth data for validation remains a limitation. Future research should focus on collecting real-time traffic sensor data or employing automated traffic counting systems to obtain ground truth data, enabling direct comparison and validation of the drone-derived results. This will provide a more comprehensive assessment of the proposed methodology's accuracy and reliability. In addition, further refinement of the detection methodology is necessary to enhance recognition and detection performance. Future work could explore more advanced models and algorithms to improve detection accuracy, particularly in complex urban environments. Incorporating simulations or conducting larger-scale comparisons with existing datasets will also help validate and optimize the model, ultimately improving its capability to address traffic-related challenges more effectively.

## ACKNOWLEDGMENTS

This study was partially supported by the Key Program of Natural Science Foundation of Inner Mongolia, China (Grant No. 2024ZD27), the Commissioned Key Program of Social Science Foundation of Inner Mongolia, China (Grant No. 2024WTZD02), and the Young Scientists Fund of the National Natural Science Foundation of China (Grant No. 61903205). The contents of this paper reflect views of the authors who are responsible for the facts and accuracy of the data presented herein. The contents of the paper do not necessarily reflect the official views or policies of the agencies.

**AUTHOR CONTRIBUTIONS**

**Qingwen Pu**: Methodology, Software, Writing – Original Draft, Visualization.
**Yuan Zhu:** Conceptualization, Writing- review & editing
**Junqing Wang:** Data and Analysis, Writing – Original Draft, Methodology, Case study
**Hong Yang:** Data, Conceptualization, Validation.
**Kun Xie:** Methodology, Conceptualization, Validation.
**Shunlai Cui：** Data Collection, Visualization

**DECLARATION OF CONFLICTING INTERESTS**

The author(s) declared no potential conflicts of interest with respect to the research, authorship, and/or publication of this article.

# REFERENCES

[1] Yang, D., K. Xie, K. Ozbay, Z. Zhao, and H. Yang. Copula-based joint modeling of crash count and conflict risk measures with accommodation of mixed count-continuous margins. *Analytic Methods in Accident Research,* Vol. 31, 2021, p. 100162.

[2] Hinton, G. E., and R. R. Salakhutdinov. Reducing the dimensionality of data with neural networks. *Science,* Vol. 313, No. 5786, 2006, pp. 504-507.

[3] LeCun, Y., Y. Bengio, and G. Hinton. Deep learning. *Nature,* Vol. 521, No. 7553, 2015, pp. 436-444.

[4] Liu, L., W. Ouyang, X. Wang, P. Fieguth, J. Chen, X. Liu, and M. Pietikäinen. Deep Learning for Generic Object Detection: A Survey. *International Journal of Computer Vision,* Vol. 128, No. 2, 2019, pp. 261-318.

[5] Krizhevsky, A., I. Sutskever, and G. E. Hinton. ImageNet classification with deep convolutional neural networks. *Communications of the ACM,* Vol. 60, No. 6, 2017, pp. 84-90.

[6] Terven, J., and D. Cordova-Esparza. A Comprehensive Review of YOLO: From YOLOv1 to YOLOv8 and Beyond. *arXiv preprint arXiv:2304.00501*, 2023.

[7] Xie, K., K. Ozbay, H. Yang, and C. Li. Mining automatically extracted vehicle trajectory data for proactive safety analytics. *Transportation research part C: emerging technologies,* Vol. 106, 2019, pp. 61-72.

[8] Yang, M., G. Han, B. Yan, W. Zhang, J. Qi, H. Lu, and D. Wang. Hybrid-SORT: Weak Cues Matter for Online Multi-Object Tracking. *arXiv preprint arXiv:2308.00783*, 2023.

[9] Zhou, Y., X. Jiang, C. Fu, and H. Liu. Operational factor analysis of the aggressive taxi speeders using random parameters Bayesian LASSO modeling approach. *Accid Anal Prev,* Vol. 157, 2021, p. 106183.

[10] Yang, D. Proactive safety monitoring: A functional approach to detect safety-related anomalies using unmanned aerial vehicle video data. 2021.

[11] Bock, J., R. Krajewski, T. Moers, S. Runde, L. Vater, and L. J. a. e.-p. Eckstein. The inD Dataset: A Drone Dataset of Naturalistic Road User Trajectories at German Intersections.In, 2019. p. arXiv:1911.07602.

[12] Bozcan, I., and E. J. a. e.-p. Kayacan. AU-AIR: A Multi-modal Unmanned Aerial Vehicle Dataset for Low Altitude Traffic Surveillance.In, 2020. p. arXiv:2001.11737.

[13] Zheng, O., M. Abdel-Aty, L. Yue, A. Abdelraouf, Z. Wang, and N. Mahmoud. CitySim: A Drone-Based Vehicle Trajectory Dataset for Safety Oriented Research and Digital Twins. *arXiv preprint arXiv:2208.11036*, 2022.

[14] Zhu, P., Y. Sun, L. Wen, Y. Feng, and Q. J. a. e.-p. Hu. Drone Based RGBT Vehicle Detection and Counting: A Challenge.In, 2020. p. arXiv:2003.02437.

[15] Geiger, A., P. Lenz, C. Stiller, and R. Urtasun. Vision meets robotics: The KITTI dataset. *The International Journal of Robotics Research,* Vol. 32, No. 11, 2013, pp. 1231-1237.

[16] Fuchs, F. B., A. R. Kosiorek, L. Sun, O. Parker Jones, and I. J. a. e.-p. Posner. End-to-end Recurrent Multi-Object Tracking and Trajectory Prediction with Relational Reasoning.In, 2019. p. arXiv:1907.12887.

[17] Guo, H., M. Keyvan-Ekbatani, and K. Xie. Lane change detection and prediction using real-world connected vehicle data. *Transportation research part C: emerging technologies,* Vol. 142, 2022, p. 103785.

[18] Xie, K., H. Yang, X. Dong, H. Yu, and H. Sun. An Automated System for Large-Scale Intersection Marking Data Collection and Condition Assessment.In, 2022.

[19] Wang, J., M. Cetin, H. Yang, K. Xie, G. Zhai, and S. Ishak. Improving Effectiveness of the Safety Service Patrol Programs: A Discrete Event-Based Simulation Approach. *Transportation Research Record,* 2024, p. 03611981241248652.

[20] Wang, C., F. Guo, R. Yu, L. Wang, and Y. Zhang. The Application of Driver Models in the Safety Assessment of Autonomous Vehicles: A Survey. *arXiv preprint arXiv:2303.14779*, 2023.

[21] Gu, X., M. Abdel-Aty, Q. Xiang, Q. Cai, J. J. A. A. Yuan, and Prevention. Utilizing UAV video data for in-depth analysis of drivers' crash risk at interchange merging areas. Vol. 123, 2019, pp. 159-169.

[22] Chen, A. Y., Y.-L. Chiu, M.-H. Hsieh, P.-W. Lin, and O. J. T. r. p. C. e. t. Angah. Conflict analytics through the vehicle safety space in mixed traffic flows using UAV image sequences. Vol. 119, 2020, p. 102744.

[23] Wu, Y., M. Abdel-Aty, O. Zheng, Q. Cai, and S. J. T. r. r. Zhang. Automated safety diagnosis based on unmanned aerial vehicle video and deep learning algorithm. Vol. 2674, No. 8, 2020, pp. 350-359.

[24] Ma, Y., H. Meng, S. Chen, J. Zhao, S. Li, and Q. J. J. o. t. e. Xiang, Part A: Systems. Predicting traffic conflicts for expressway diverging areas using vehicle trajectory data. Vol. 146, No. 3, 2020, p. 04020003.

[25] Chen, X., Z. Li, Y. Yang, L. Qi, and R. J. I. T. o. I. T. S. Ke. High-resolution vehicle trajectory extraction and denoising from aerial videos. Vol. 22, No. 5, 2020, pp. 3190-3202.

[26] Klautau, A., I. Correa, F. Bastos, I. Nascimento, J. Borges, A. Oliveira, P. Batista, and S. Lins. Integrated simulation of deep learning, computer vision and physical layer of UAV and ground vehicle networks.In *Deep Learning and Its Applications for Vehicle Networks*, CRC Press. pp. 321-342.

[27] Wang, Y., Z. Huang, R. Laganière, H. Zhang, and L. Ding. A UAV to UAV tracking benchmark. *Knowledge-Based Systems,* Vol. 261, 2023, p. 110197.

[28] Di Capua, M., A. Ciaramella, and A. De Prisco. Machine Learning and Computer Vision for the automation of processes in advanced logistics: the Integrated Logistic Platform (ILP) 4.0. *Procedia Computer Science,* Vol. 217, 2023, pp. 326-338.

[29] Zhang, N., H. Liu, and K. H. Low. UAV Collision Risk Assessment in Terminal Restricted Area by Heatmap Representation.In *AIAA SCITECH 2023 Forum*, 2023. p. 0737.

[30] Gazzea, M., A. Miraki, O. Alisan, M. M. Kuglitsch, I. Pelivan, E. E. Ozguven, and R. Arghandeh. Traffic monitoring system design considering multi-hazard disaster risks. *Scientific reports,* Vol. 13, No. 1, 2023, p. 4883.

[31] Xing, Z., S. Yang, X. Zan, X. Dong, Y. Yao, Z. Liu, and X. Zhang. Flood vulnerability assessment of urban buildings based on integrating high-resolution remote sensing and street view images. *Sustainable Cities and Society,* Vol. 92, 2023, p. 104467.

[32] Alharbi, A., I. Petrunin, and D. Panagiotakopoulos. Deep Learning Architecture for UAV Traffic-Density Prediction. *Drones,* Vol. 7, No. 2, 2023, p. 78.

[33] Westhofen, L., C. Neurohr, T. Koopmann, M. Butz, B. Schütt, F. Utesch, B. Neurohr, C. Gutenkunst, and E. Böde. Criticality metrics for automated driving: A review and suitability analysis of the state of the art. *Archives of Computational Methods in Engineering,* Vol. 30, No. 1, 2023, pp. 1-35.

[34] Tzutalin. Tzutalin. LabelImg. Git code (2015). https://github.com/tzutalin/labelImg. 2015.

[35] Ren, S., K. He, R. Girshick, J. J. I. T. o. P. A. Sun, and M. Intelligence. Faster R-CNN: Towards Real-Time Object Detection with Region Proposal Networks. Vol. 39, No. 6, 2017, pp. 1137-1149.

[36] Liu, W., D. Anguelov, D. Erhan, C. Szegedy, S. Reed, C.-Y. Fu, and A. C. Berg. SSD: Single Shot MultiBox Detector.In, Springer International Publishing, Cham, 2016. pp. 21-37.

[37] Redmon, J., and A. J. a. e.-p. Farhadi. YOLOv3: An Incremental Improvement. 2018.

[38] Bochkovskiy, A., C. Y. Wang, and H. Liao. YOLOv4: Optimal Speed and Accuracy of Object Detection. 2020.

[39] Gupta, H., O. P. J. M. T. Verma, and Applications. Monitoring and surveillance of urban road traffic using low altitude drone images: a deep learning approach. pp. 1-21.

[40] Tian, J., Q. Jin, Y. Wang, J. Yang, S. Zhang, and D. Sun. Performance analysis of deep learning-based object detection algorithms on COCO benchmark: a comparative study. *Journal of Engineering and Applied Science,* Vol. 71, No. 1, 2024, p. 76.

[41] Antwi, R. B., S. Takyi, A. Karaer, E. E. Ozguven, R. Moses, M. A. Dulebenets, and T. Sando. Detecting School Zones on Florida's Public Roadways Using Aerial Images and Artificial Intelligence (AI2). *Transportation Research Record,* Vol. 2678, No. 4, 2024, pp. 622-636.

[42] Hao, W., and S. Zhili. Improved mosaic: Algorithms for more complex images.In *Journal of Physics: Conference Series, No. 1684*, IOP Publishing, 2020. p. 012094.

[43] Gao, O., C. Niu, W. Liu, T. Li, H. Zhang, and Q. Hu. E-DeepLabV3+: A Landslide Detection Method for Remote Sensing Images.In *2022 IEEE 10th Joint International Information Technology and Artificial Intelligence Conference (ITAIC), No. 10*, IEEE, 2022. pp. 573-577.

[44] Hou, X., Y. Wang, and L.-P. Chau. Vehicle tracking using deep sort with low confidence track filtering.In *2019 16th IEEE International Conference on Advanced Video and Signal Based Surveillance (AVSS)*, IEEE, 2019. pp. 1-6.

[45] Zhang, Y., P. Sun, Y. Jiang, D. Yu, F. Weng, Z. Yuan, P. Luo, W. Liu, and X. Wang. Bytetrack: Multi-object tracking by associating every detection box.In *European Conference on Computer Vision*, Springer, 2022. pp. 1-21.

[46] Du, Y., Z. Zhao, Y. Song, Y. Zhao, F. Su, T. Gong, and H. Meng. Strongsort: Make deepsort great again. *IEEE Transactions on Multimedia*, 2023.

[47] Khan, Z., S. M. Khan, K. Dey, and M. Chowdhury. Development and evaluation of recurrent neural network-based models for hourly traffic volume and annual average daily traffic prediction. *Transportation Research Record,* Vol. 2673, No. 7, 2019, pp. 489-503.

[48] Tian, Y., and L. Pan. Predicting short-term traffic flow by long short-term memory recurrent neural network.In *2015 IEEE international conference on smart city/SocialCom/SustainCom (SmartCity)*, IEEE, 2015. pp. 153-158.

[49] Lane, V. M. Obstacle detection and tracking in an urban environment using 3d lidar and a mobileye 560.In, Massachusetts Institute of Technology, 2017.

[50] Vasudevan, V., R. Agarwala, and A. Tiwari. LiDAR-Based Vehicle–Pedestrian Interaction Study on Midblock Crossing Using Trajectory-Based Modified Post-Encroachment Time. *Transportation Research Record*, 2022, p. 03611981221083295.

[51] Fu, C., and H. Liu. Investigating influence factors of traffic violations at signalized intersections using data gathered from traffic enforcement camera. *PLoS one,* Vol. 15, No. 3, 2020, p. e0229653.